\documentclass[]{jfm}
\usepackage{graphicx}
\usepackage{dcolumn}
\usepackage{bm}
\usepackage{subfigure}
\usepackage{natbib}
\usepackage{epstopdf}
\usepackage{amsmath}
\usepackage{graphicx}
\usepackage{dcolumn}
\usepackage{amssymb,amsmath}
\usepackage{natbib}
\usepackage{subfigure}
\usepackage{epstopdf}
\usepackage{placeins}
\usepackage{citesort}
\usepackage{color}
\usepackage{epstopdf}

\def\eps{{\epsilon}}

\title[Unifying theory of scaling in thermal convection]{The unifying theory of scaling in thermal convection: The updated prefactors} 

\author[Stevens, van der Poel, Grossmann, Lohse]{Richard J. A. M. Stevens $^{1,2}$, Erwin P. van der Poel $^{2}$, Siegfried Grossmann$^{3}$, and Detlef Lohse$^{2}$}%
\affiliation{
$^1$Dept. of Mechanical Engineering, Johns Hopkins University, Baltimore, Maryland 21218, USA\\
$^2$Physics of Fluids Group, Faculty of Science and Technology,  J.M. Burgers Center for Fluid Dynamics, and  MESA+ Institute, University of Twente, 7500 AE  Enschede, The Netherlands\\
$^3$ Fachbereich Physik, Philipps-Universit\"{a}t Marburg, Am Renthof 6, D-35032 Marburg, Germany\\
}
  
\date{\today}

\begin{document}
\maketitle

\begin{abstract}
The unifying theory of scaling in thermal convection (\cite{gro00}) (henceforth the GL theory) suggests that there are no pure power laws for the Nusselt and Reynolds numbers as function of the Rayleigh and Prandtl numbers in the experimentally accessible parameter regime. In \cite{gro01} the dimensionless parameters of the theory were fitted to 155 experimental data points by \cite{ahl01} in the regime $3\times 10^7 \le Ra \le 3 \times 10^{9}$ and $4\le Pr \le 34$ and \cite{gro02} used the experimental data point from \cite{qiu01b} and the fact that $Nu(Ra,Pr)$ is independent of the parameter $a$, which relates the dimensionless kinetic boundary thickness with the square root of the wind Reynolds number, to fix the Reynolds number dependence. Meanwhile the theory is on one hand well confirmed through various new experiments and numerical simulations. On the other hand these new data points provide the basis for an updated fit in a much larger parameter space. Here we pick four well established (and sufficiently distant) Nu(Ra,Pr) data points and  show that the resulting $Nu(Ra,Pr)$ function is in agreement with almost all established experimental and numerical data up to the ultimate regime of thermal convection, whose onset also follows from the theory. One extra $Re(Ra,Pr)$ data point is used to fix $Re(Ra,Pr)$. As $Re$ can depend on the definition and the aspect ratio the transformation properties of the GL equations are discussed in order to show how the GL coefficients can easily be adapted to new Reynolds number data while keeping $Nu(Ra,Pr)$ unchanged. 
\end{abstract}

\section{Introduction}\label{sec:intro}
Thermal convection is omnipresent in science and technology and its paradigmatical representation is Rayleigh-B\'enard (RB) convection: a fluid in a sample heated from below and cooled from above. This system has received considerable attention in the last decades \citep{ahl09,sig94,loh10}, with one focus on the scaling properties of the global heat transport of the system. The now widely  accepted viewpoint is the Grossmann-Lohse (GL) theory \citep{gro00,gro01,gro02,gro04}. The basis for this theory of scaling in RB convection are exact global balances for the energy and thermal dissipation rates derived from the Boussinesq equations and the decomposition of the flow in boundary layer (BL) and bulk contributions. The scaling of the dissipation rates in the BLs is assumed to obey Prandtl-Blasius-Pohlhausen scaling \citep{sch79}, which is justified as long as the shear-Reynolds numbers of the BLs are not too large, and the scaling relations in the bulk are estimated based on Kolmogorov-type arguments for homogeneous isotropic turbulence.  While the theory gives the different {\it scaling relations} for the individual contributions to the energy dissipation rates in the bulk and in the BL, namely $\eps_{u,bulk}$ and $\eps_{u,BL}$, and to the thermal dissipation rates in the bulk (background) and in the BLs (plus the plumes, see \cite{gro04}), namely $\eps_{\theta,bulk}$ and $\eps_{\theta,BL}$, the {\it absolute sizes} of these four relative contributions are not given by the theory. They are expressed in four dimensionless prefactors $c_i$, $i=1,2,3,4$ for $\eps_{u,BL}$, $\eps_{u,bulk}$,  $\eps_{\theta,BL}$, and  $\eps_{\theta,bulk}$, respectively, which have to be obtained from experimental or numerical data for $Nu(Ra,Pr)$.

When the theory was developed early this century, such data were scarce and often contradicting each other, due to sidewall and plate effects, insufficient knowledge of the material properties of the fluid, lack of numerical resolution and other problems. \cite{gro01} used 155 data points for $Nu(Ra,Pr)$ in the parameter range $3\times 10^7 \le Ra \le 3 \times 10^{9}$ and $4\le Pr \le 34$ obtained by \cite{ahl01}, which was the most extensive data set at that time. This fixed $Nu(Ra,Pr)$ for {\it all} $Ra$ and $Pr$, considered as valid up to the meanwhile found (\cite{he11}) ultimate regime of thermal convection, where the Prandtl-Blasius type BL becomes unstable. $Re(Ra,Pr)$ was fixed (cf. \cite{gro02}) with one extra adoption of the prefactor $a$, i.e. the amplitude parameter of the Prandtl BL thickness, in the Prandtl-Blasius scaling relation $\lambda_u = a L / \sqrt{Re}$, to the experimental data of \cite{qiu01b}, where $\lambda_u$ is the mean thickness of the kinetic BL and $L$ the height of the sample.

\begin{figure}
\begin{center}
\subfigure{\includegraphics[width=0.80\textwidth]{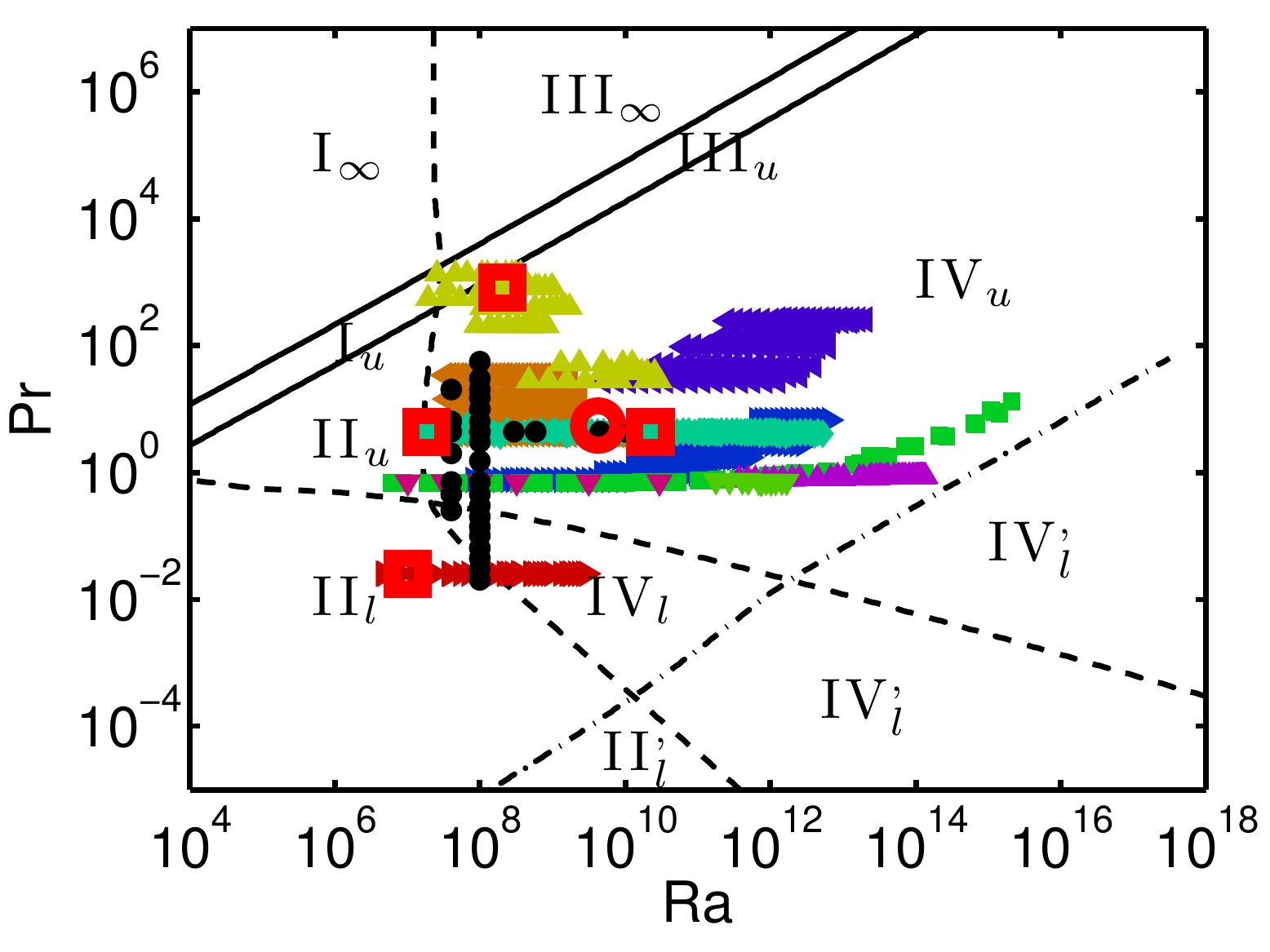}}
\subfigure{\includegraphics[width=0.99\textwidth]{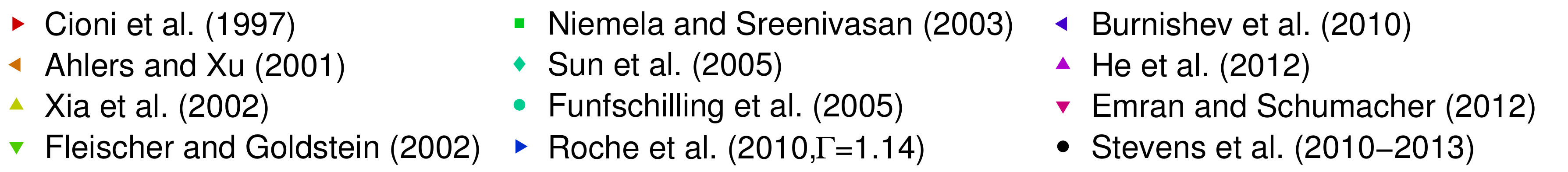}}
    \caption{Phase diagram in $Ra-Pr$ plane for RB convection according to the GL-model (\cite{gro00,gro01,gro02,gro04}) in a $\Gamma=1$ sample with no-slip boundary conditions. The upper solid line means $Re=1$; the lower nearly parallel solid line corresponds to $\epsilon_{u,BL}=\epsilon_{u,bulk}$; the curved solid and dashed line is $\epsilon_{\theta,BL}=\epsilon_{\theta,bulk}$; and along the long-dashed line $\lambda_u=\lambda_\theta$, i.e., $2aNu = \sqrt{Re}$. The dash-dotted line indicates where the laminar kinetic BL is expected to become turbulent, based on a critical shear Reynolds number $Re_s^*=1014$ of the kinetic BL, see text. The data are from \cite{cio97,gla99,ahl01,fle02,xia02,cha02,nie03,sun05e,fun05,roc10,bur10,emr12,he12b,ste10a,ste10b,poe13}. Note that for the Stevens et al.\ data, points from different papers have been combined in the graph. The four large red open squares indicate the location of the four $Nu(Ra,Pr)$ points and the large red open circle the $Re(Ra,Pr)$ point that have been used for the new GL-fit.}
\label{fig:figure1}
\end{center}
\end{figure}

Although the data to which we adopted the four prefactors $c_i$ and $a$ were relatively {\it local} in parameter space, the theory was rather successful in describing the {\it global} behavior  $Nu(Ra,Pr)$ and also $Re(Ra,Pr)$, as described in detail in \cite{ahl09}. This included the prediction that for $Pr \approx 1$ the onset to the ultimate regime should take place when Ra is of the order of $10^{14}$. This prediction was based on an assumed onset of a sheared BL instability at a shear Reynolds number $Re_s \approx 420$, which is the value given in \cite{ll87}. Indeed, very recently \cite{he11} have found the onset of the ultimate regime at this very Rayleigh number. Thanks to joint efforts of the community the experimental and numerical data situation for $Nu(Ra,Pr)$ has considerably improved in the last decade. Measurements have been extended to a much larger domain in the Ra-Pr parameter space, see the updated phase diagrams in figure \ref{fig:figure1} and figure \ref{fig:figure8},  and plate- and sidewall corrections are much better understood and taken into account (\cite{bro05,ahl00,roc01b,ver02,nie03,ahl09}). One notices that for $\Gamma=1/2$ higher Ra number values have been obtained than for $\Gamma=1$, while the $Pr$ number dependence is much more explored for $\Gamma=1$ than for $\Gamma=1/2$. Furthermore, due to the increasing computational power and better codes the numerical data are now well converged, confirming and complementing the experimental data. Meanwhile \cite{ste10,ste10d} achieved $Ra= 2 \cdot 10^{12}$ at $Pr= 0.7$ in a $\Gamma = 1/2$ sample and obtained a good agreement with the experimental data of \cite{he11} and \cite{nie00}. 

This situation calls for a {\it refit} of the four prefactors $c_i$ and $a$ of the GL theory, in spite of the success of the theory with the coefficients of \cite{gro01}:  It is clear that the surface  $Nu(Ra, Pr)$ above the Ra-Pr parameter space will be much more stable and "wobble'' less if we put it on four distant and trustable "legs'' $Nu_i(Ra_i,Pr_i)$, $i=1,2,3,4$, rather than putting it on four "legs'' somewhere in the center but close to each other. As we will see $Nu(Ra,Pr)$ is only determined by the choice of these four $Nu(Ra,Pr)$ data points from experiments. The accuracy of the GL-fit is verified by comparing it with several data sets over a wide parameter regime and by making a second fit that reveals in which regimes there is some uncertainty. Including more data points in the fitting procedure does not lead to better fits since data tend to be clustered in the phase space. Therefore including more data points increases the weight of some data without actually adding additional physical information. An additional Reynolds number measurement is necessary to fix $a$ and the relation between $a$ and $Re_L$, which is the Reynolds number for which no bulk is left and the whole flow consists of laminar boundary layer as will be explained below. The shortcoming of the old set of $c_i$, $a$, and $Re_L$ was particularly obvious for small $Pr$, say $Pr \le 1$ (see figure \ref{fig:figure5}), because at the days of \cite{gro01} no reliable information was available in that parameter regime and therefore no Nusselt data of that regime had been included into the fit.

The structure of the paper is as follows: In section 2 we will provide the refit of the GL theory for an aspect ratio  $\Gamma = 1$, leading to  $Nu(Ra,Pr)$ in the whole parameter space up to the ultimate state. In section 3 we discuss the robustness of the fit. In section 4 we will show that this fit also describes the available data for $\Gamma = 1/2$ and will in particular discuss the onset of the ultimate regime. Section 5 gives conclusions and an outlook on the new challenges.

\section{Refit of the GL theory for $\Gamma = 1$} \label{refit}

\begin{figure}
\subfigure{\includegraphics[width=0.49\textwidth]{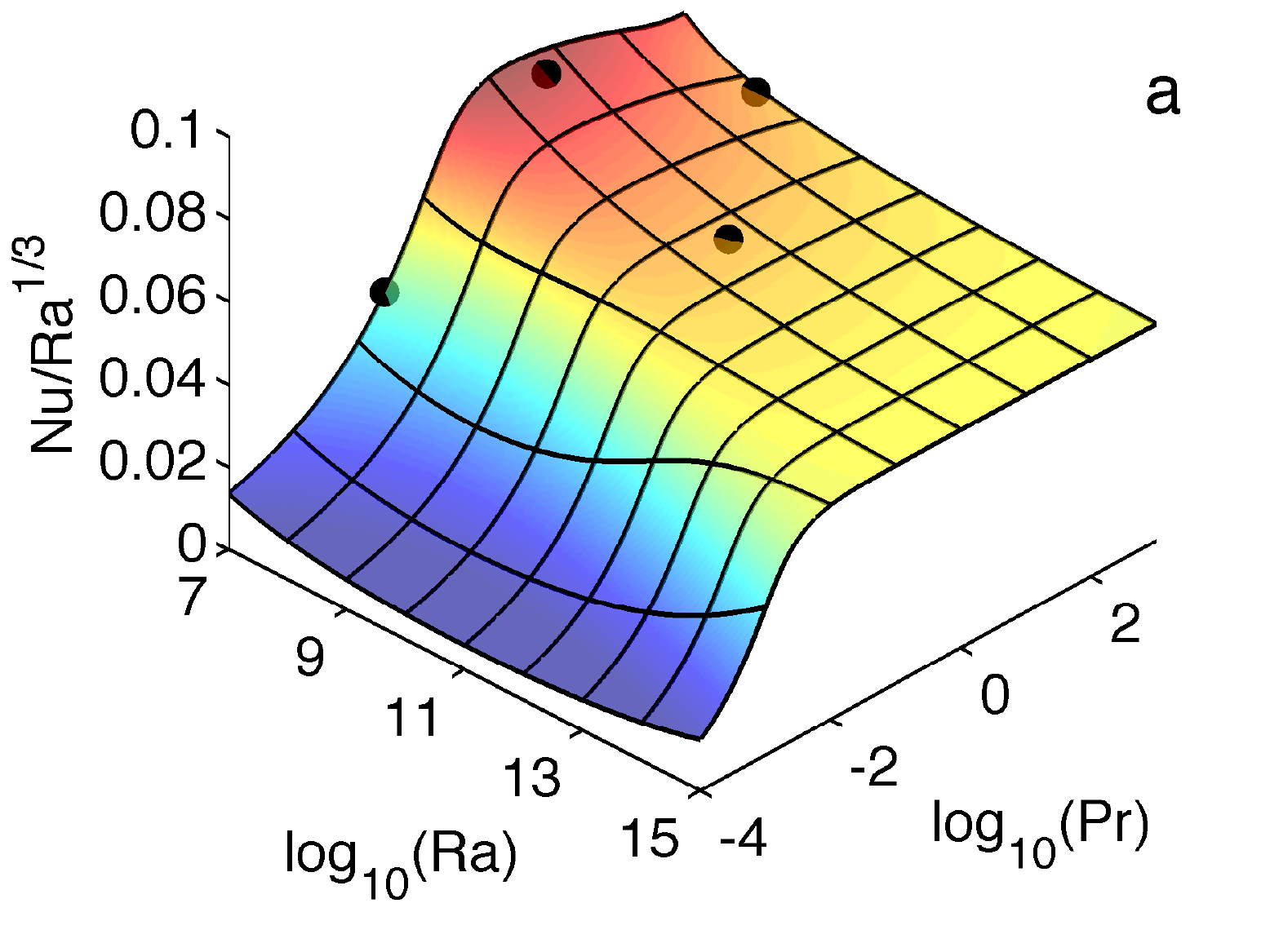}}
\subfigure{\includegraphics[width=0.49\textwidth]{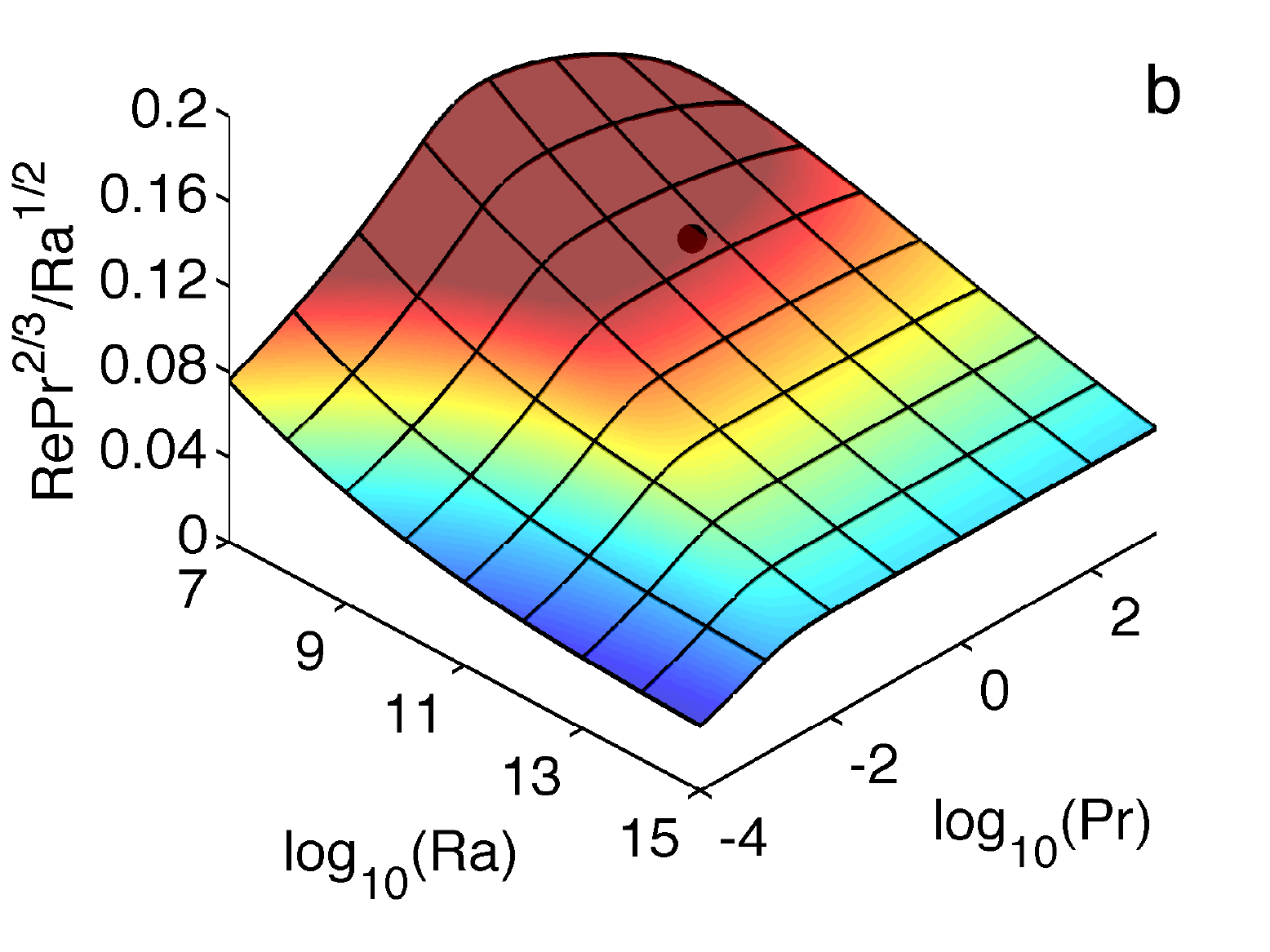}}
    \caption{Compensated three-dimensional visualization of a) Nu(Ra,Pr) and b) Re(Ra,Pr). The four $Nu(Ra,Pr)$ points and the $Re(Ra,Pr)$ point used to fit the GL parameters $c_1=8.05$, $c_2=1.38$,  $c_3=0.487$, $c_4=0.0252$ have been indicated by the black points in the Nu(Ra,Pr) and Re(Ra,Pr) graph, respectively.}
\label{fig:figure2}
\end{figure}

The GL theory describes Nu(Ra,Pr) and Re(Ra,Pr) with the following two coupled equations (\cite{ahl09}),
\begin{eqnarray} \label{eq1}
(Nu-1)Ra Pr^{-2}&=& c_1 \frac{Re^2}{g(\sqrt{Re_L/Re})}+c_2Re^3,\\
\label{eq2}
Nu-1 &=&c_3 Re^{1/2}Pr^{1/2} \left\{ f \left[ \frac{2a Nu}{\sqrt{Re_L}} g \left(\sqrt{\frac{Re_L}{Re}} \right) \right] \right\}^{1/2} + \nonumber \\
          &   &c_4 Pr Re f \left[ \frac{2a Nu}{\sqrt{Re_L}} g \left(\sqrt{\frac{Re_L}{Re}} \right)\right],
\end{eqnarray}
where the crossover functions f and g model the crossover from the thermal boundary layer nested in the kinetic one towards the inverse situation and that for which $\lambda_u\sim L$ looses its scaling with Re since $\lambda_u$ extends to sample half-height $L/2$ and cannot increase further with decreasing Re; for details see \cite{gro01}. As described by \cite{gro02} the prefactor $a$ has to be determined from experimental data. Whereas the definition of the Nusselt number is very clear there are various reasonable ways to define a Reynolds number. We decided to use one experimental data point of \cite{qiu01b} to determine the value of $a$ and from figure 1 of \cite{gro02} we read Ra=4.2$\cdot 10^9$, Pr=5.5, Re=2.1 $\cdot 10^3$. In addition we demand for the Reynolds number $Re_L$ that $\lambda_u=aL/\sqrt{Re_L}=L/2$, meaning that $Re_L=(2a)^2$ is fixed for given $a$.

\begin{figure}
\centering
\subfigure{\includegraphics[width=0.49\textwidth]{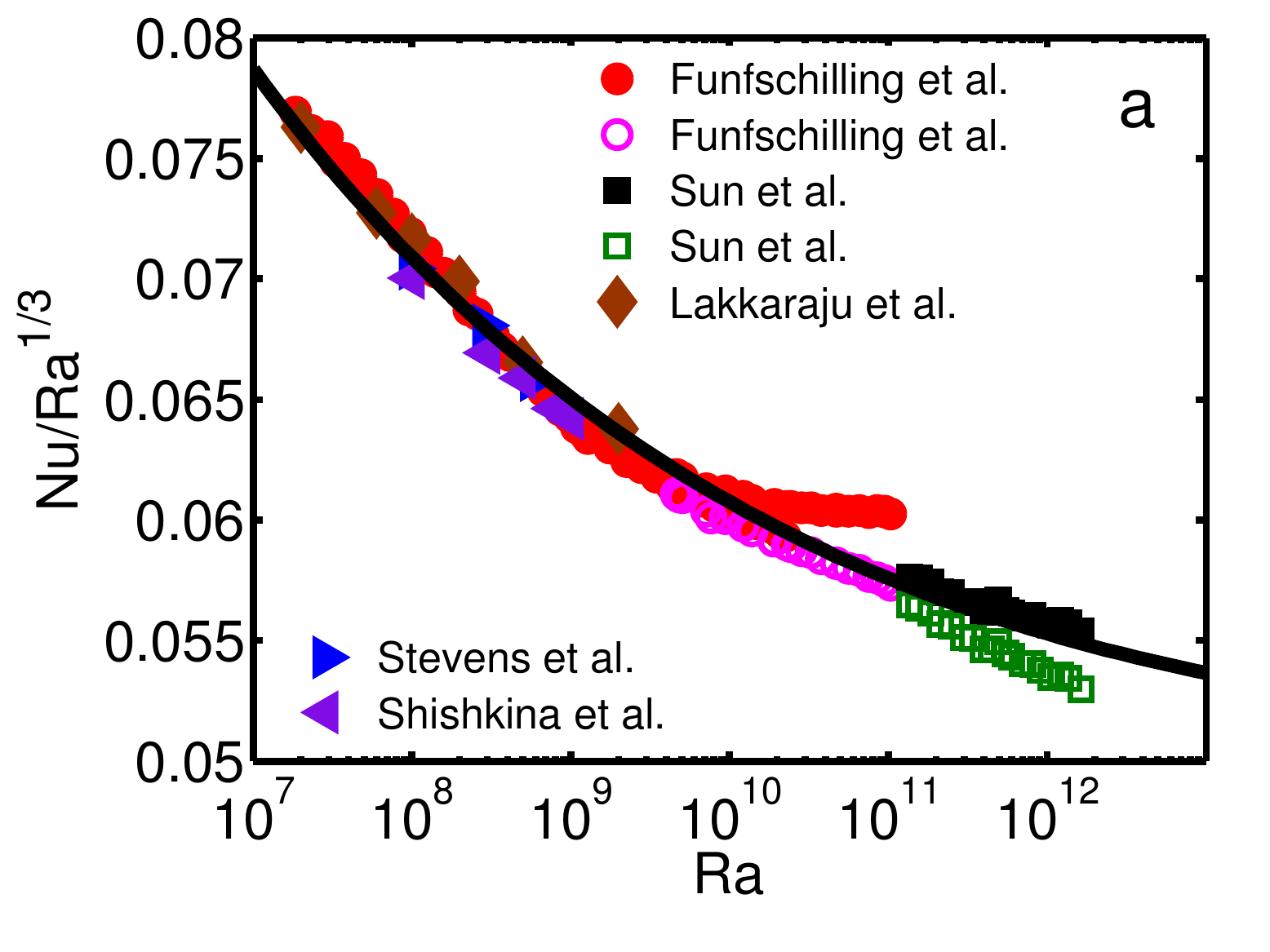}}
\subfigure{\includegraphics[width=0.49\textwidth]{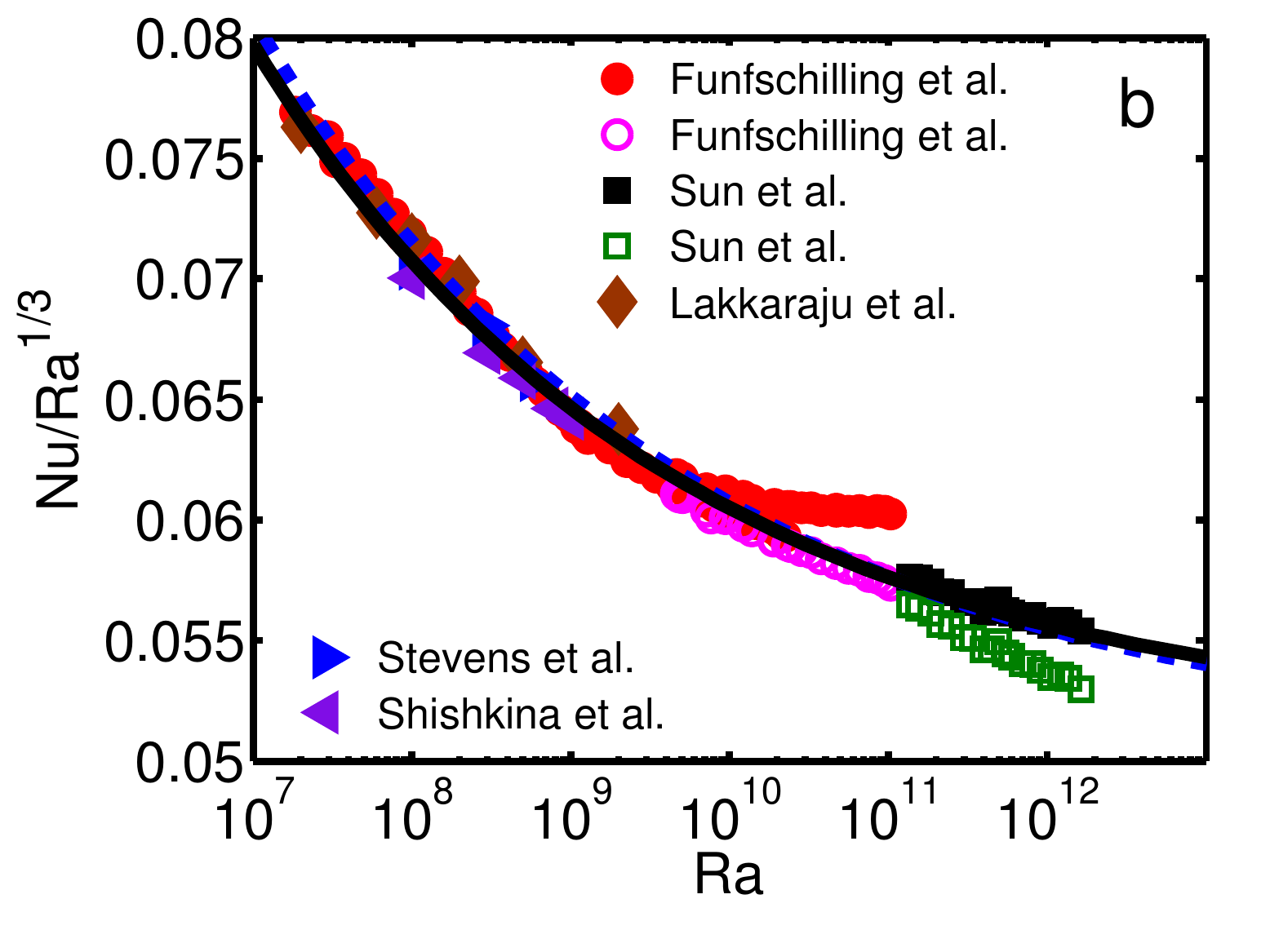}}
\caption{Comparison of the $Ra$-scaling of the original GL-fit from \cite{gro01} (a) with the new fit (b) for water, i.e. $Pr=4.38$ unless mentioned otherwise, in a $\Gamma=1$ sample. The circles (\cite{fun05}) and squares (\cite{sun05d}) indicate experimental results. Open symbols indicate the uncorrected data and solid symbols the data after correction for the finite plate conductivity. The diamonds (\cite{lak12}, $Pr=5.4$), right pointing triangles (\cite{ste11b}), and left pointing triangles (\cite{shi09}) indicate results from numerical simulations. The solid (black) line indicate the GL fit of section \ref{refit} and the dashed (blue) line the GL fit in section \ref{Robustness}.}
\label{fig:figure3}
\end{figure}

\begin{figure}
\centering
\subfigure{\includegraphics[width=0.49\textwidth]{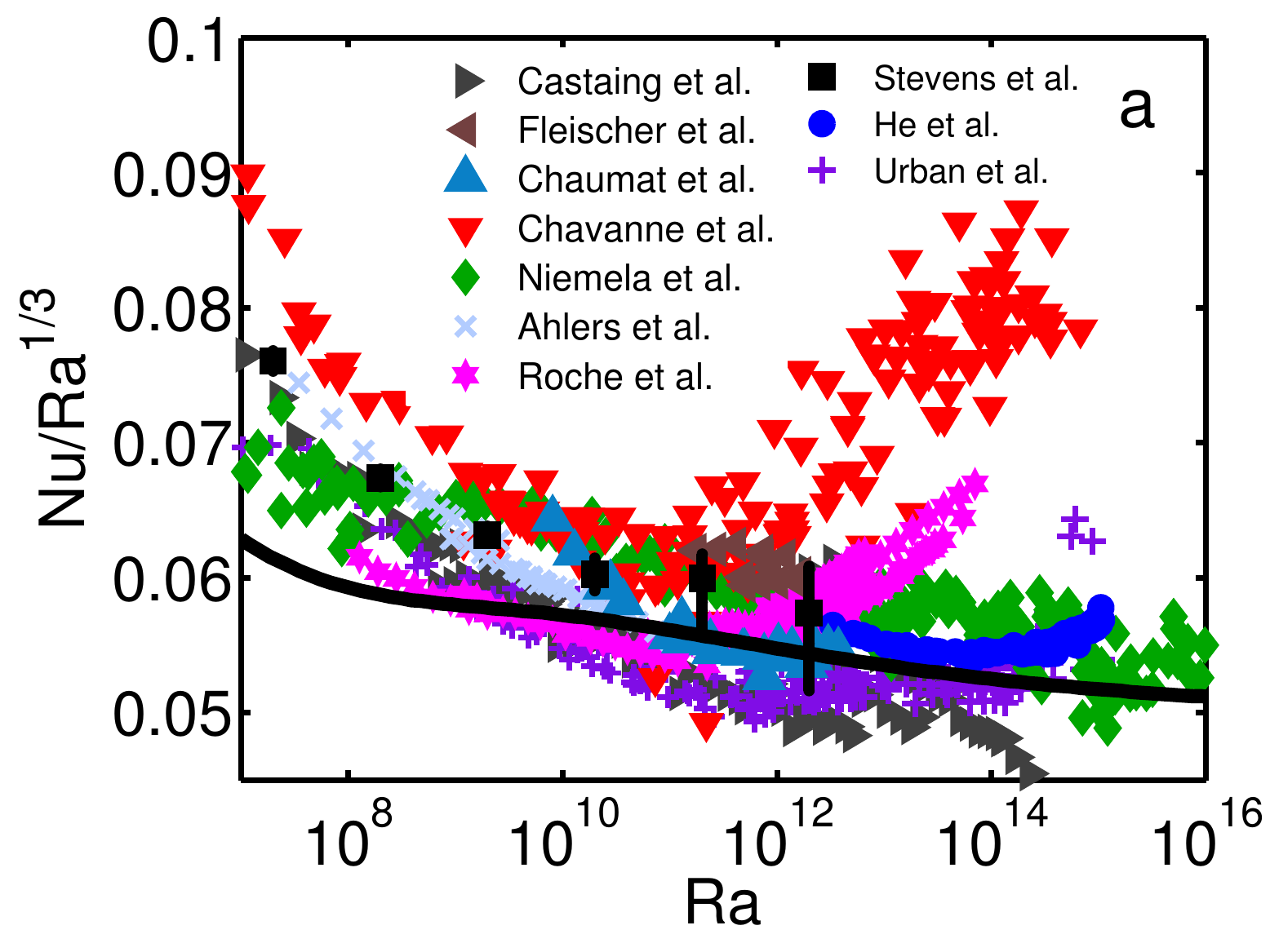}}
\subfigure{\includegraphics[width=0.49\textwidth]{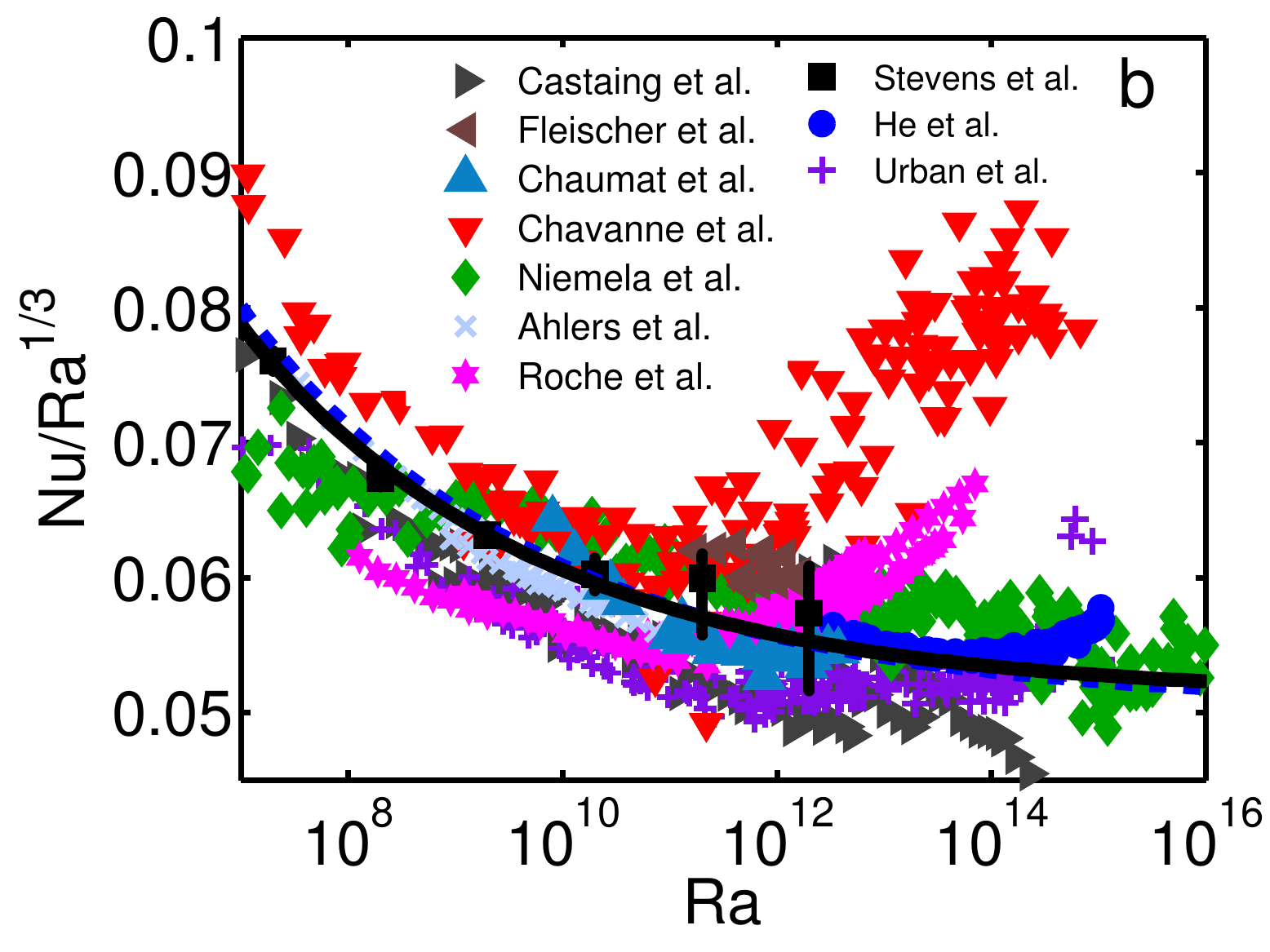}}
\caption{Comparison of the $Ra$-scaling of the original GL-fit from \cite{gro01} (a) with the new fit (b) in a $\Gamma=1/2$ sample and varying $Pr$, see phase diagram in figure \ref{fig:figure8}. The right pointing triangles are the experimental data from \cite{cas89} with wall corrections \cite{roc10}, left pointing triangles (\cite{fle02}), upward pointing triangles (\cite{cha02}), downward pointing triangles (\cite{cha01}), diamonds (\cite{nie00}), crosses (\cite{ahl09c}), hexagons (\cite{roc10}), circles (\cite{he11,ahl12b}), and plusses (\cite{urb11,urb12}) indicate experimental data and the squares results from numerical simulations (\cite{ste10,ste10d}). The solid (black) line indicate the GL fit of section \ref{refit} and the dashed (blue) line the GL fit in section \ref{Robustness}.}
\label{fig:figure4}
\end{figure}

\begin{figure}
\centering
\subfigure{\includegraphics[width=0.49\textwidth]{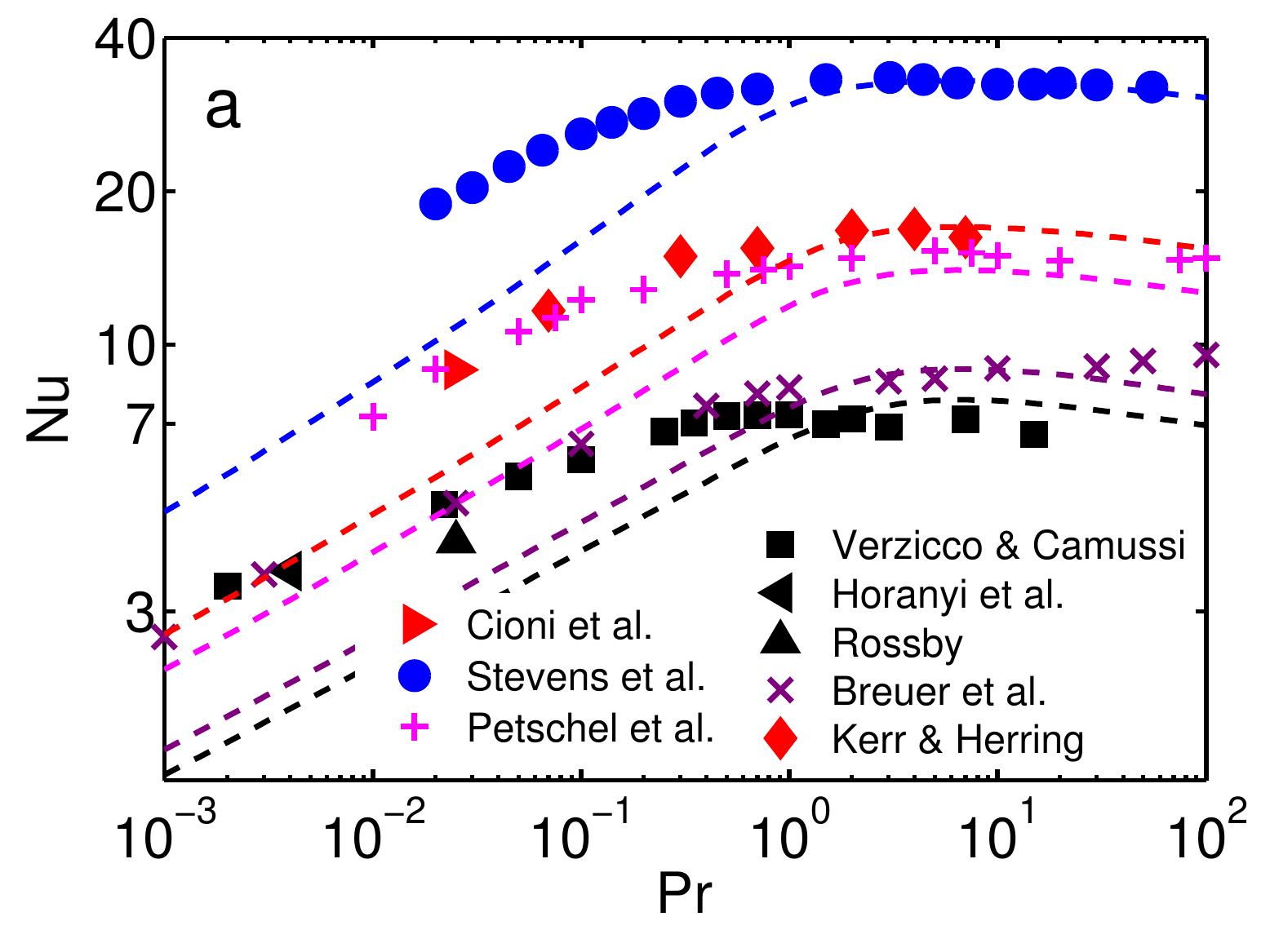}}
\subfigure{\includegraphics[width=0.49\textwidth]{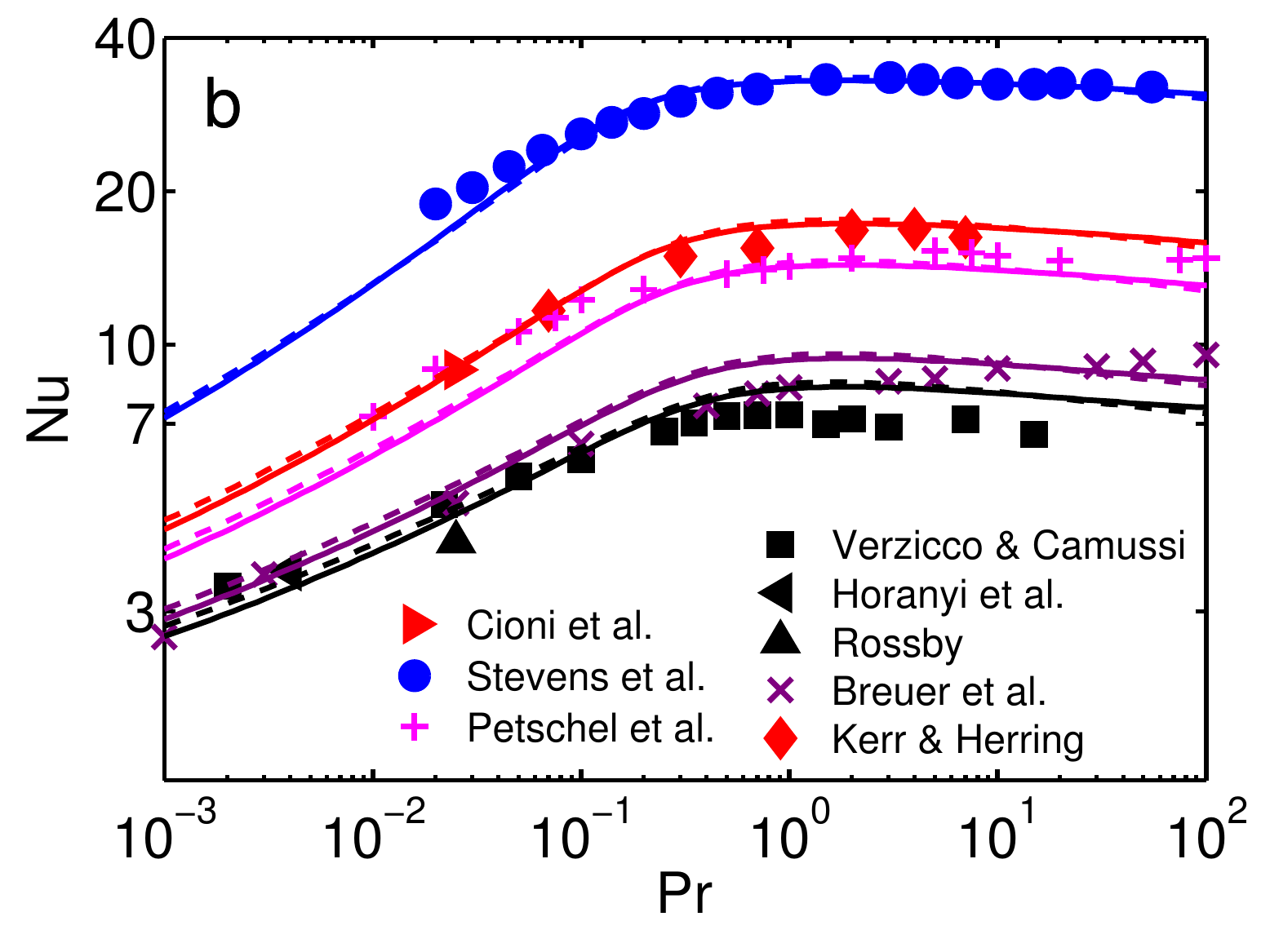}}
\caption{Comparison of the $Pr$-scaling of the original GL-fit from \cite{gro01} (a) with the new fit (b) for different $Ra$. The right pointing triangle (\cite{cio97}), left pointing triangle (\cite{hor99}), and the upward pointing triangle (\cite{ros69}) indicate experimental results. The circles (\cite{poe13}) and squares (\cite{ver99}) indicate numerical results obtained in a cylinder with aspect ratio $\Gamma=1$. The plusses (\cite{pet13}) indicate numerical results obtained in a periodic domain and the diamonds (\cite{ker00}) and crosses (\cite{bre04}) numerical results obtained in a box with free slip boundary condition at the sidewall. The colors black, purple, magenta, red, and blue (from bottom to top) corresponds to the $Ra$ numbers $5\times10^5$, $10^6$, $5\times10^6$, $10^7$, and $10^8$. The solid lines indicate the GL fit of section \ref{refit} and the dashed lines the GL fit in section \ref{Robustness}.}
\label{fig:figure5}
\end{figure}

In order to get accurate values for the four dimensionless prefactors $c_i$, it is necessary to choose four data points with as much information on the richness of the RB system as possible, which means that data points from different regimes should be selected. Therefore we determined the $c_i$ from the data points of \cite{fun05} at $Ra=1.8\times10^7$ and $Ra=2.25\times10^{10}$, both with $Pr=4.38$, the data point from \cite{xia02} with $Pr=818$ at $Ra=2.04\times10^8$, and the data point from \cite{cio97} at $Ra=1\times10^7$ with $Pr=0.025$. The location of these data points in the RB phase diagram is indicated by the large red squares in figure \ref{fig:figure1} and by the black dots in the corresponding three-dimensional Nu(Ra,Pr) visualization in figure \ref{fig:figure2}a. Figure \ref{fig:figure1} shows that these data are indeed within different regimes. The reason for choosing these specific data points is two-fold. First of all we consider these four data points to be reliable. And apart from the data point by \cite{xia02}, which is the only experiment in that large $Pr$ regime, all data points agree very well with experimental or numerical data from other groups, see figures \ref{fig:figure3} and \ref{fig:figure5}. In addition, these four data points are relatively far apart in the Ra-Pr parameter space to ensure that they provide the theory with as much information on the richness of the RB physics as possible. To provide information on the $Ra$-scaling we selected the measurements of \cite{fun05} at $Ra=1.8\times10^7$ and $Ra=2.25\times10^{10}$ with $Pr=4.38$. In order to include information on the transition between the 'upper' and 'lower' regimes, which is modeled by the crossover functions $f$ and $g$, it is necessary to include data points in the low, intermediate, and high Pr number regime. We do this selecting next to the intermediate Pr number data from \cite{fun05}, the low $Pr=0.025$ number measurement by \cite{cio97} at $Ra=1\times10^7$ and the high $Pr=818$ measurement by \cite{xia02} at $Ra=2.04\times10^8$. Altogether the four data points provide information from three different Pr numbers and four different Ra numbers. From these four data points, and an initial guess for $a$, we determine the $c_i$ with a fourth order Newton-Raphson root finding method or by using a trust-region-reflective optimization algorithm, which both gave the same result. Subsequently, the $Re(Ra,Pr)$ point of \cite{qiu01b} is used to find the appropriate value of $a$ with the transformation property of the GL model (\cite{gro02}), which is described below in detail. Due to this transformation property of the GL equations the four $Nu(Ra,Pr)$ data points determine the Nusselt number dependence, while the Re number data point of \cite{qiu01b} fixes the absolute value of the Reynolds number throughout the phase space. This results in the following five GL parameters $c_1=8.05$, $c_2=1.38$,  $c_3=0.487$, $c_4=0.0252$, and $a=0.922$. The difference in significant number is due to the fact that some coefficients are less sensitive to uncertainty than others. 

It was pointed out by \cite{gro02} that $Nu(Ra,Pr)$ is invariant and thus independent of the parameter $a$ under the following transformation
\begin{eqnarray} \label{eq1}
	Re	&\rightarrow & \alpha Re,\\ \label{trans1}
	a	& \rightarrow & \alpha^{1/2} a,\\ \label{trans2}
	c_1	& \rightarrow & c_1/\alpha^2,\\ \label{trans3}
	c_2	& \rightarrow & c_2/\alpha^3,\\ \label{trans4}
	c_3	& \rightarrow & c_3/\alpha^{1/2},\\ \label{trans5}
	c_4	& \rightarrow & c_4/\alpha,\\ \label{trans6}
	Re_L&\rightarrow & \alpha Re_L. \label{trans7}
\end{eqnarray}
We note that the above transformation is consistent with the relation $Re_L=(2a)^2$ used above and allows for the transformation of the above set of coefficients to different Reynolds number definitions or aspect ratios. Such a new set of coefficients is obtained by first determining $\alpha$. Here $\alpha$ is determined as $Re_1(Ra,Pr)/Re_2(Ra,Pr)$, where $Re_1$ is the Reynolds number value of a measurement point in the data set at a given $Ra$ and $Pr$ and $Re_2$ is the Reynolds number value obtained from the GL model with the coefficients mentioned above. Subsequently, equations (\ref{trans1}) to (\ref{trans7}) can be used to calculate the new coefficients.

In figures \ref{fig:figure3} to \ref{fig:figure5} we compare the original GL-fit from \cite{gro01} with this new GL-fit. These figures clearly reveal that the new GL-fit is much closer to the data in the low $Pr$ number regime, while maintaining the similar excellent agreement for the high $Pr$ number data as before. We emphasize that this excellent agreement with all other presently available data from experiments and simulations confirms that the $c_i$ and $a$ values we calculated describes $Nu(Ra,Pr)$ well in the regime that is nowadays covered by state of the art experiments and simulations. It is also noteworthy that  figure \ref{fig:figure3} and \ref{fig:figure4} show that the $Nu$ number scaling with $Ra$ is well predicted by the GL-theory for $Ra$ values that are decades higher than the highest $Ra$ number point that is used to determine the $c_i$ values, namely $Ra=2.25\times10^{10}$, thus showing the predictive power of the GL-theory.

\section{Robustness} \label{Robustness}
To illustrate the robustness of the fit presented above, we made a second fit to four other data points, i.e.\ the data points from \cite{fun05} at $Ra=2.96\times10^7$ and $Ra=1.92\times10^{10}$ with $Pr=4.38$, the one from \cite{xia02} at $Ra=2.24\times10^8$ with $Pr=554$ and finally the data point by \cite{ker00} at $Ra=10^7$ with $Pr=0.07$. Three out of these four data points lie relatively close to the original four data points, but the low $Pr = 0.07$ point from \cite{ker00} substantially differs from the original $Pr = 0.025$. The reason that three of the four points are close to the original four points in the $Ra-Pr$ parameter space is that one can only select "reliable legs" in regimes were many measurements have been done and these regimes only cover a limited part of the parameter space.

\begin{figure}
\centering
\subfigure{\includegraphics[width=0.49\textwidth]{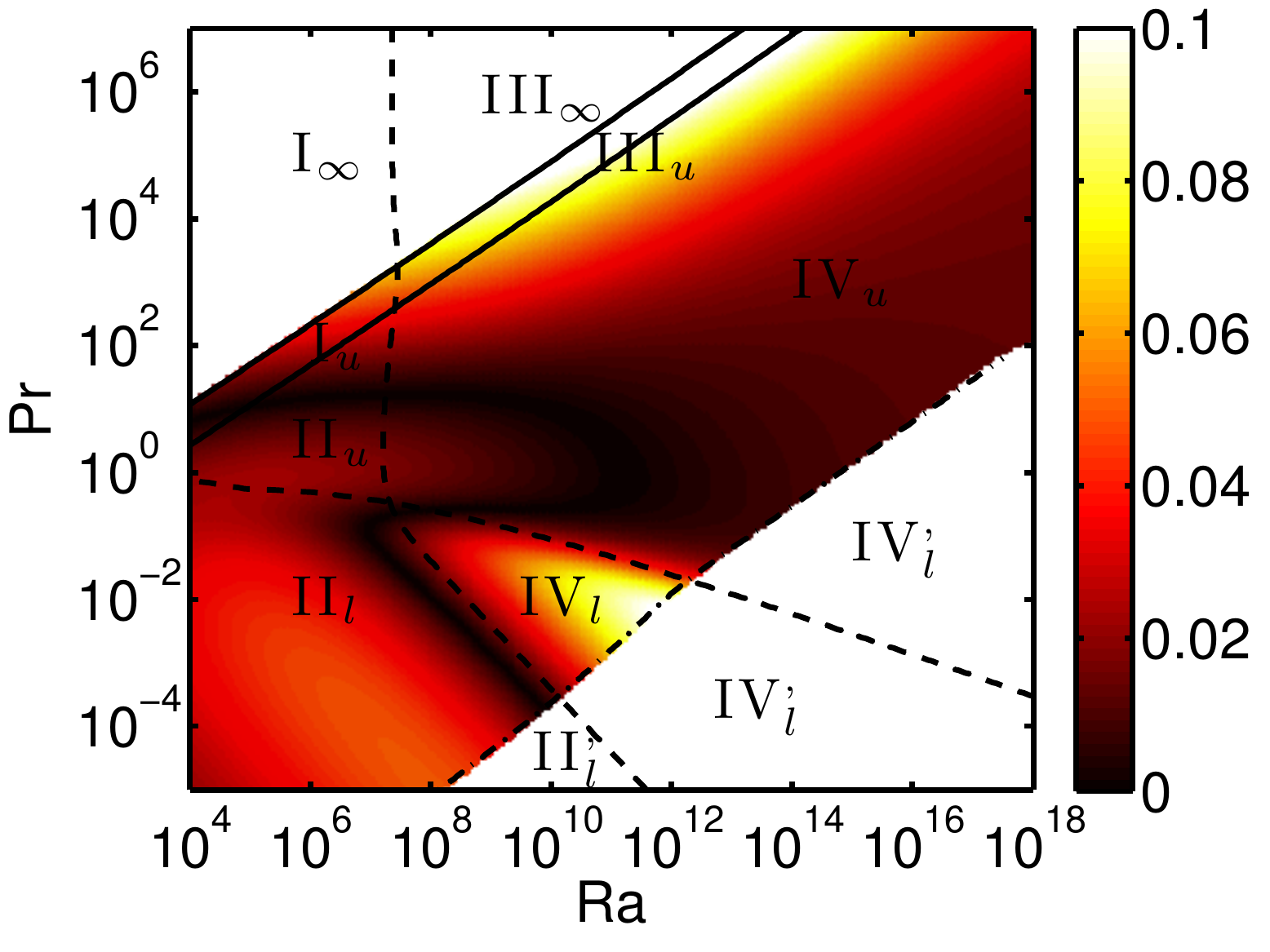}}
    \caption{Relative difference between $Nu$ calculated from the original fit and $Nu$ calculated from the additional fit. The color scale ranges from black to white, indicating agreement mostly up to $\approx 4\%$. Only in two small ranges, $III_u$ and $IV_l$, it goes up to $10\%$.}
\label{fig:figure6}
\end{figure}

\begin{figure}
\centering
\subfigure{\includegraphics[width=0.49\textwidth]{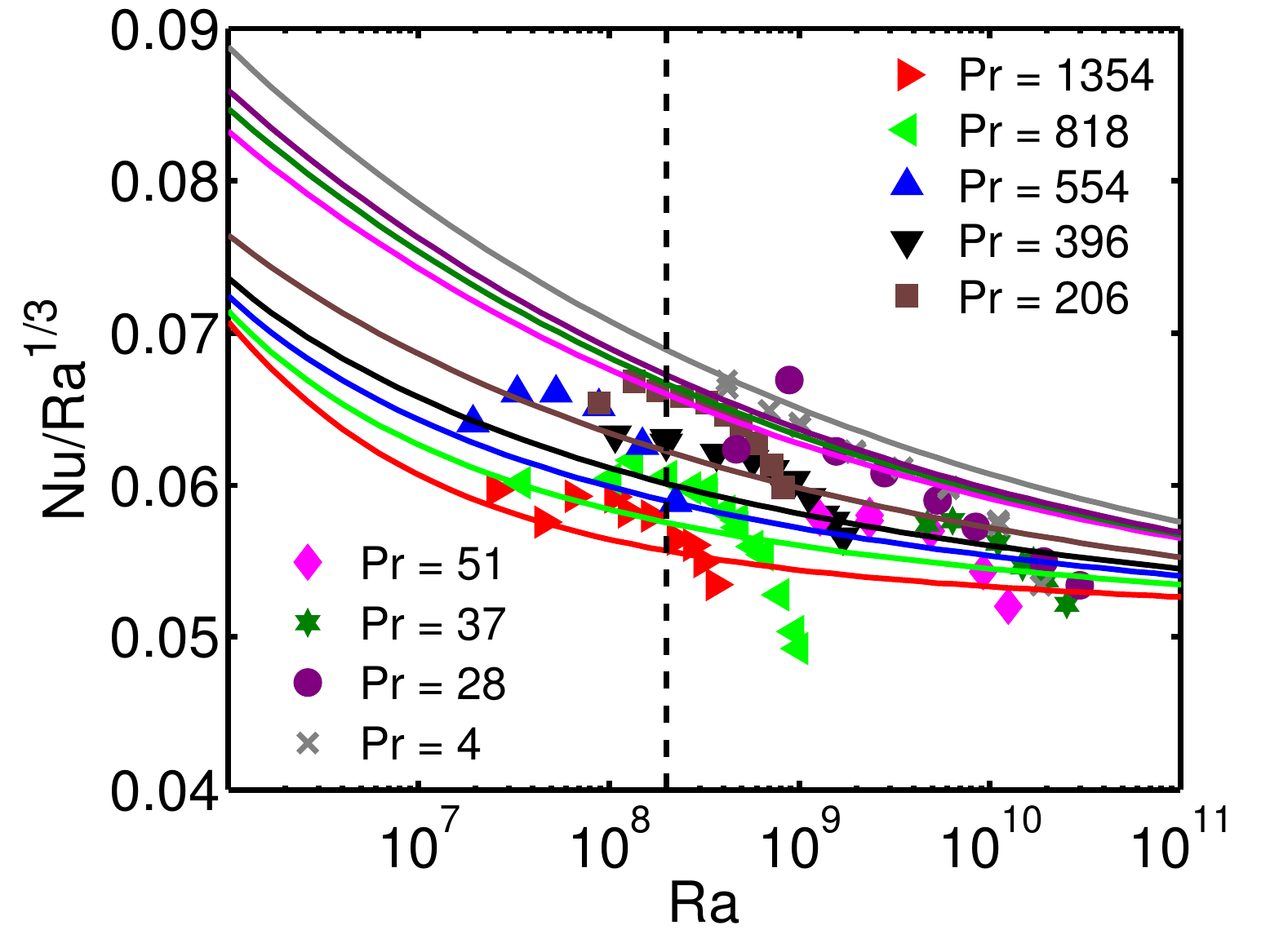}}
\subfigure{\includegraphics[width=0.49\textwidth]{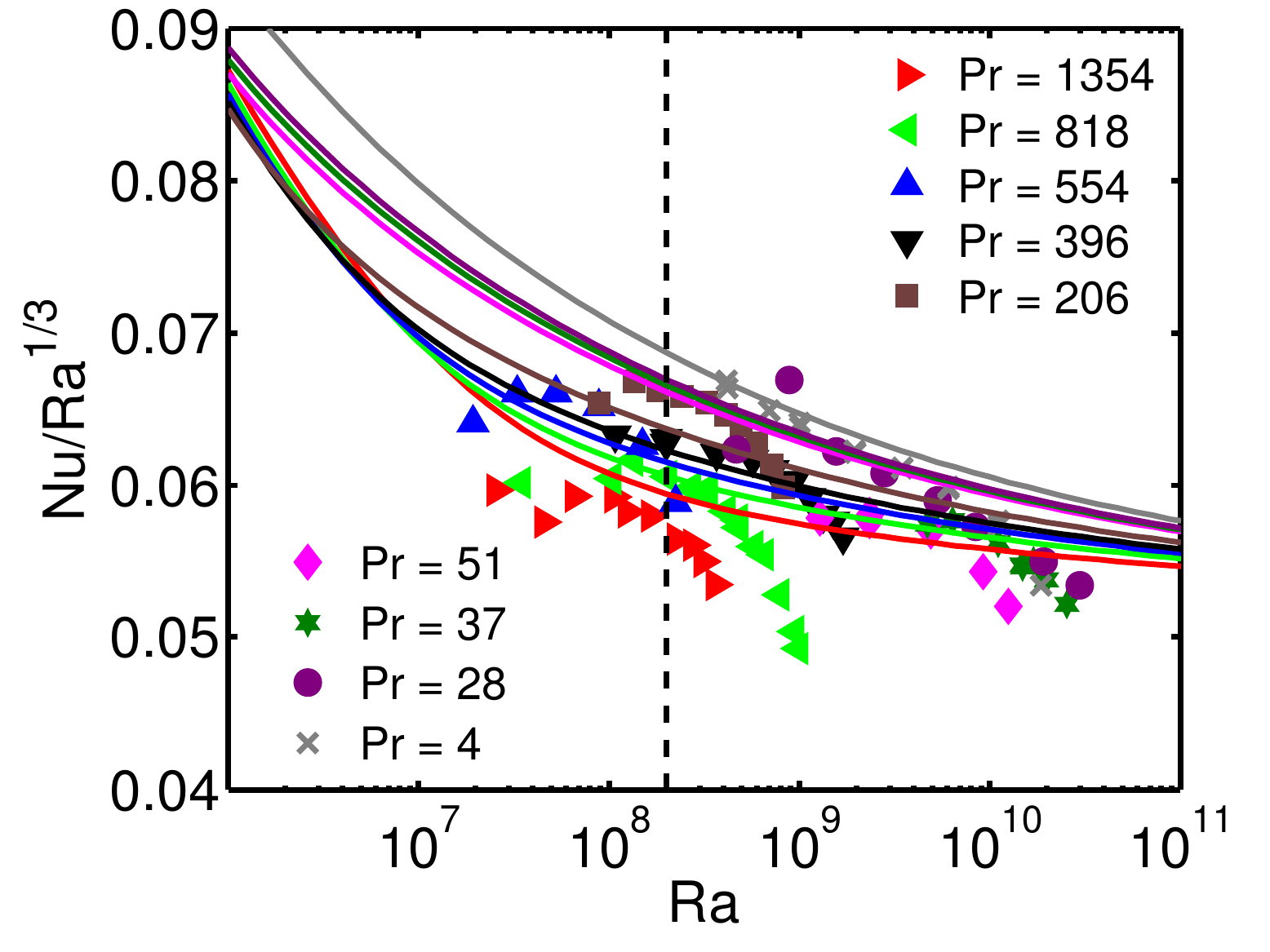}}
\subfigure{\includegraphics[width=0.49\textwidth]{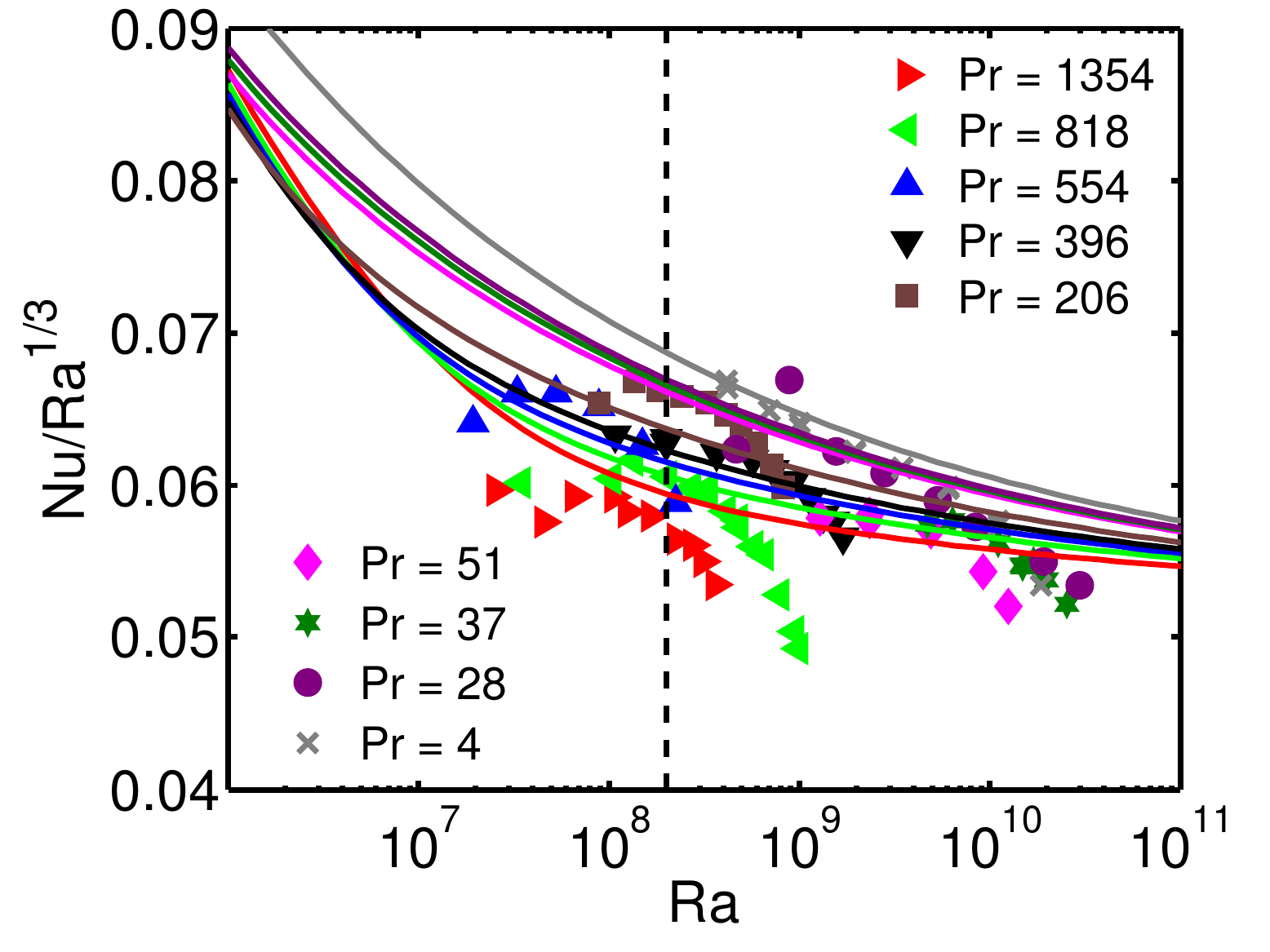}}
\subfigure{\includegraphics[width=0.49\textwidth]{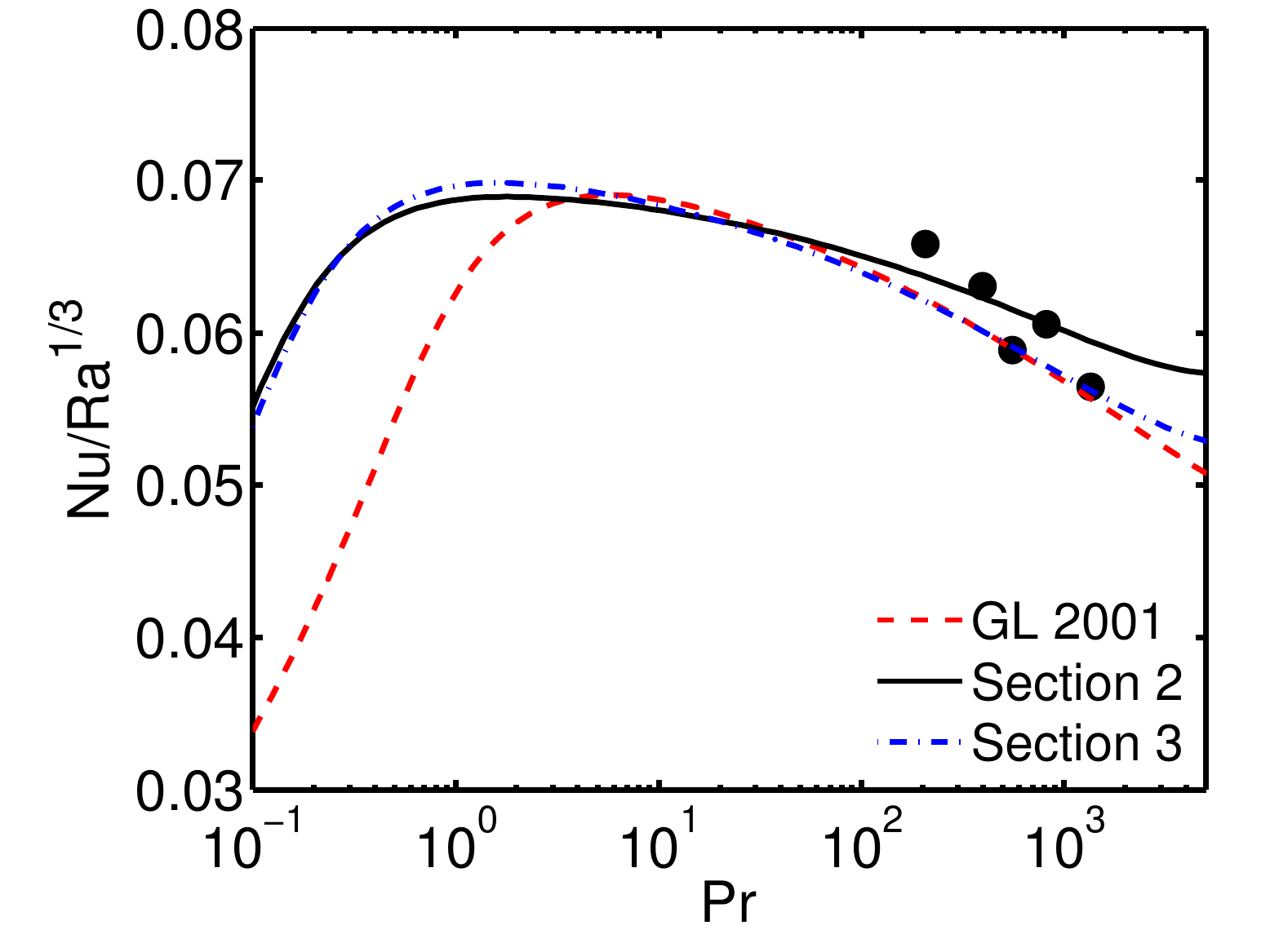}}
    \caption{Panel a, b, and c show the Nusselt number measurements of \cite{xia02} versus $Ra$ in comparison to the original GL-fit from \cite{gro01} (panel a), the GL fit presented in section \ref{refit} (panel b), and the second fit discussed in section \ref{Robustness}. Panel d shows the $Pr$ number dependence of the \cite{xia02} experiments for $Ra\approx2\times10^8$ compared to the three different GL fits mentioned above.}
\label{fig:figure7}
\end{figure}

The resulting GL coefficients are $c_1=11.8$, $c_2= 1.33$, $c_3= 0.528$, $c_4= 0.0222$, and $a = 0.843$ compared to $c_1=8.05$, $c_2=1.38$, $c_3=0.487$, $c_4=0.0252$, and $a=0.922$ of the fit described above. In order to compare the two fits we show both fits together with experimental and numerical data from several experiments in figures \ref{fig:figure3}, \ref{fig:figure4}, \ref{fig:figure5}, and \ref{fig:figure7}. In addition we give the relative difference in $Nu(Ra,Pr)$ calculated in the fit described in the previous section and $Nu$ calculated from this additional fit in the parts of the parameter space where the GL fit is valid in figure \ref{fig:figure6}. A comparison between both fits shows that the difference is very minor in the regimes $IV_u$, $II_u$, and $I_u$, and that the differences increase in the regimes $II_l$, $IV_l$, and $III_u$, which are very far away from the region in the parameter space where reliable data points are available. The reason is that a very small variation in the measurements point can lead to significant differences if the implied information is extrapolated over many decades in $Ra$ and $Pr$ using the GL-theory. For the fits compared here the differences increase up to about $10\%$. We find that the differences are mainly caused by the uncertainty in the \cite{xia02} data, which is reflected by the two different data points we took from this data set. In figure \ref{fig:figure7} we compare the \cite{xia02} measurements with the original GL-fit from \cite{gro01}, and the two fits presented in this work. Panels a, b and c show that the measured $Nu/Ra^{1/3}$ decreases faster with increasing Ra than predicted by the GL model. A comparison of the \cite{xia02} data with other data obtained at $Pr=4.38$ shows that the measurements in the lower $Ra$ number regime collapse very well with other measurements, while for the higher $Ra$ the measured Nusselt number seems a little lower in the \cite{xia02} experiments than in other experiments. Figure \ref{fig:figure7}d compares the \cite{xia02} measurements between $Ra\approx2\times10^8$ and $Ra\approx2.4\cdot$ $10^8$ with the different GL-fits and shows that the fit presented in section \ref{refit} uses a high Pr number point that aligns with the reliable $Pr=4.38$ data point, while the second fit  uses a high $Pr$ number point of the lower branch. In this way the uncertainty of these measurements is reflected by the two fits and as is shown in figure \ref{fig:figure6} this difference is mostly visible near regime $I_\infty$ and $III_\infty$.

\section{GL theory for $\Gamma = 1/2$ and ultimate regime}
In principle,  the $c_i$ depend on the aspect ratio $\Gamma$. However, it is well known that only small differences in $Nu$ are observed between $\Gamma=1/2$ and $\Gamma=1$ (\cite{ahl09}). This weak aspect ratio dependence is confirmed by figure \ref{fig:figure4}, which shows that the $Ra$ number scaling for $Pr=0.7$ in a $\Gamma=1/2$ sample is captured very accurately by the new fit for $\Gamma=1$, and in the low Ra number regime the new fit is even much better than the original GL-fit from \cite{gro01}. 

\begin{figure}
\begin{center}
\subfigure{\includegraphics[width=0.80\textwidth]{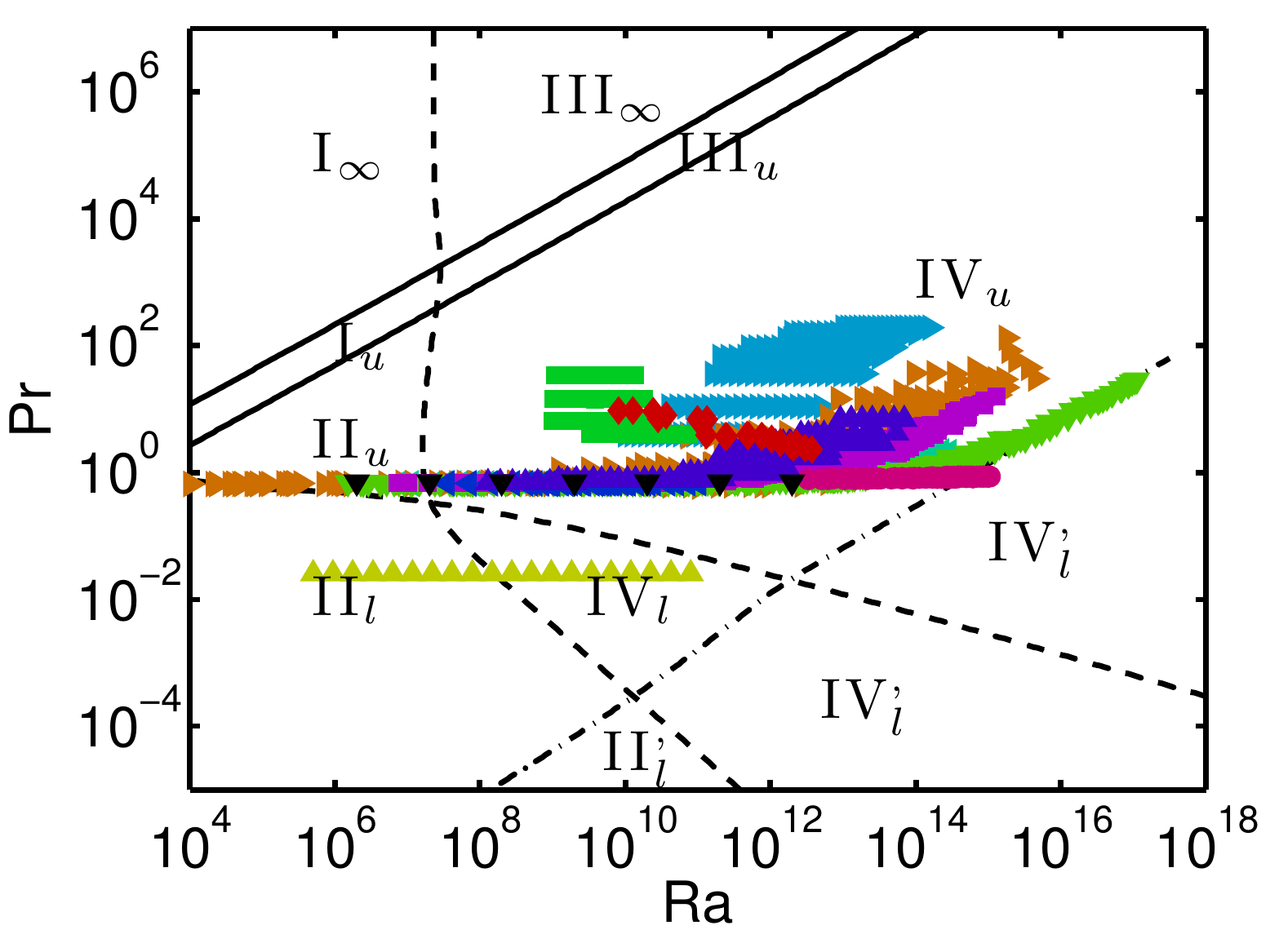}}
\subfigure{\includegraphics[width=0.99\textwidth]{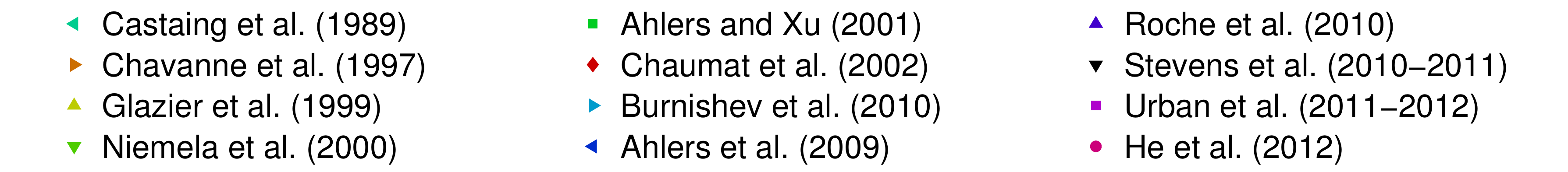}}
    \caption{Phase diagram in $Ra-Pr$ plane for RB convection in a $\Gamma=1/2$ sample with no-slip boundary conditions. The lines are the same as in figure \ref{fig:figure1}. The data are from \cite{cas89,cha97,gla99,nie00,ahl01,cha02,ahl09,bur10,roc10,urb11,urb12,he11,ste10,ste10d}.} 
\label{fig:figure8}
\end{center}
\end{figure}

The location in Ra-Pr space of the various regimes of the GL theory is based on the coefficients $c_i$ and $a$. The updated lines that encompass the regimes are plotted in the phase diagrams shown in figures \ref{fig:figure1} and \ref{fig:figure8}. The line that indicates the onset of the ultimate regime, where the kinetic boundary layer has become turbulent, is now based on the new coefficients, the transition at $Ra = 5\cdot10^{14}$, observed by \cite{he11} for $Pr = 0.86$, and the $Re$ number measurements by \cite{qiu01b}. This gives $Re_s^*=1039$ ($a=0.922$) and for the second fit we made, see section \ref{Robustness}, we get  $Re_s^*=954$ ($a = 0.843$). In a $\Gamma=1/2$ sample \cite{he11} found experimentally that $Re=0.252Ra^{0.434}Pr^{0.750}$ using a recently developed and tested elliptic approximation (\cite{he06,zho09d,he10,he11b,zho11b}), which defines $Re$ unambiguously, based on properties of correlation functions. Using this relation at $Ra=10^{13}$ and $Pr=0.86$ this gives $Re_s^*=572$ with $a=0.684$ for the first fit, see section \ref{refit}, and $Re_s^*=521$ with $a=0.623$ for the second fit we made, see section \ref{Robustness}. These $Re_s^*$ values are different from the previously used $Re_s^* = 420$ with $a=1.72$ taken from pipe flow (\cite{ll87}). From the transformation property of the GL equations one gets that $Re_s^*$ increases by a factor of $\alpha$ when $Re$ increases by a factor $\alpha$, while $a$ increases by a factor $\sqrt{\alpha}$. The $a$ values found here are significantly different from $a=0.482$, which was found by \cite{gro02}. Calculating the $Re_s^*$ values equivalent to $a=0.482$ from the $a$ and $Re_s$ combinations mentioned above, i.e.\ $a = 0.843$ with $Re_s^*=954$, $a=0.922$ with $Re_s^*=1039$, $a=0.684$ with $Re_s^*=572$, and $a=0.623$ with $Re_s^*=521$, gives  $Re_s^*=298 \pm 15$ for the new coefficients, so we see that the notification of $Re_s^*$ alone, without $a$ is not sufficient.

The phase diagram in figure \ref{fig:figure8} shows that the measurements of \cite{he11} up to $Ra \approx 10^{15}$ at $Pr = 0.86$ are up to now the only experiments that have reached the ultimate regime. They observe the onset of the ultimate regime at $Ra = 5\cdot10^{14}$ and a transition region for $10^{13} \leq Ra \leq 5\cdot 10^{14}$. The experiments by \cite{he11} are the only room temperature experiments for $Ra\gtrsim10^{12}$, while all other experiments that have reached these Ra numbers are low temperature experiments with Helium close to the critical point (\cite{cha97,cha01,nie00,nie01,nie06,roc10,urb11,urb12}). In these low temperature experiments it is difficult to reach the ultimate regime because the Pr number increases with increasing Ra, see figure \ref{fig:figure8}. Nevertheless the low temperature experiments by \cite{nie00} seem to come very close to the ultimate regime and one may wonder why the transition region observed by \cite{he11} was  not observed in the \cite{nie00} experiments. As discussed in detail by \cite{ahl12b}, presumably the scatter of the \cite{nie00} data at this highest Ra (which seems to be due primarily to the uncertainties in the fluid properties) as well as the fact that the transition is smooth are the reasons for this. Figure \ref{fig:figure4} shows that the magnitude of the scatter in the \cite{nie00} data is similar to the observed increase in the compensated Nusselt number in the transition regime by \cite{he11}. The phase diagram also shows that other low temperature experiments by \cite{cha97,cha01}, \cite{roc10}, and \cite{urb11,urb12} do not reach the ultimate regime and therefore no transition to the ultimate regime due to a boundary layer shear instability is expected in these experiments. 
 
\section{Conclusions and outlook}
In this paper we have used the availability of new experimental and numerical data, and our increased understanding of the physics of the Rayleigh-B\'enard system to determine the prefactors of the unifying theory for scaling in thermal convection, i.e. the GL theory, much more accurately. The resulting $Nu(Ra,Pr)$ function is in very good agreement with almost all established experimental and numerical data up to the ultimate regime of thermal convection, and has significantly improved the predictions. In figure \ref{fig:figure4} one can notice the onset of the ultimate regime in the Nu(Ra) scaling of the measurements of \cite{he11}. Extensions of the GL theory to the ultimate regime by \cite{gro11} are able to explain the observed Reynolds number scaling in that regime as well as the origin of the log-profiles observed in the ultimate regime by \cite{ahl12}.

In line with \cite{gro01},  we have determined the prefactors from experimental measurements. This has great value as it shows that the information of only five data-points is sufficient to accurately predict $Nu(Ra,Pr)$ and $Re(Ra,Pr)$ up to the ultimate regime. All is based on the GL theory, which builds on exact global balances for the energy and thermal dissipation rates, derived from the Boussinesq equations, and the decomposition of the flow in boundary layer and bulk contributions. A finding with further implications is that the value $a$, i.e.\ the amplitude parameter of the Prandtl BL thickness, is higher than found by \cite{gro02}. This $a$ value is for example used by \cite{shi10} to determine the number of grid points that should be placed in the boundary layers. \cite{shi10} compared the theoretical predictions with results from simulations in $\Gamma=1/2$ samples. For this aspect ratio the newly found value of {a ($a=0.684$ for fit of section \ref{refit} and $a=0.623$ of section \ref{Robustness}) is higher, but still relatively close to the previously used $a=0.482$. However, it looks like that for $\Gamma=1$ case $a$ is even higher ($a=0.922$ for the fit of section \ref{refit} and $a=0.843$ for the fit of section \ref{Robustness}), which could have implications for the resolution that should be used in simulations. This finding confirms the conclusions of \cite{ste10} who pointed out that the only way to really confirm that the used numerical resolution is sufficient is to obtain the same Nusselt number with different grids resolutions as there is namely always some uncertainty in estimates of the required grid resolution.

A further challenge we want to pursue is to calculate the $c_i$ and $a$ directly from the fluid equations, without the input of any experimental or numerical data, or at least quantitatively relate their values to important fluid concepts like Prandtl-Blasius-Pohlhausen theory, the von Karman-Prandtl theory, etc.\ in order to get an even deeper understanding of the GL theory.

\section*{Acknowledgments}
We thank all of our colleagues for various discussions over the years and G.\ Ahlers, F.\ Chilla, K.\ Petschel, P.\ Roche, L.\ Skrbek, R.\ Verzicco, and K.-Q.\ Xia for providing us with their experimental and numerical data and for numerous scientific discussions and insightful remarks on RB over the years. Special thanks goes to G.\ Ahlers for his insightful comments on earlier versions of this manuscript. We acknowledge the Foundation for Fundamental Research of Matter (FOM), which is part of NWO, for funding.

\end{document}